# A Rough Computing based Performance Evaluation Approach for Educational Institutions

Debi Prasanna Acharjya and Debarati Bhattacharjee

*School of Computing Sciences and Engineering, VIT University, Vellore, India*

*dpacharjya@gmail.com, bhattacharjee.debarati@gmail.com*

## Abstract

*Performance evaluation of various organizations especially educational institutions is a very important area of research and needs to be cultivated more. In this paper, we propose a performance evaluation for educational institutions using rough set on fuzzy approximation spaces with ordering rules and information entropy. In order to measure the performance of educational institutions, we construct an evaluation index system. Rough set on fuzzy approximation spaces with ordering is applied to explore the evaluation index data of each level. Furthermore, the concept of information entropy is used to determine the weighting coefficients of evaluation indexes. Also, we find the most important indexes that influence the weighting coefficients. The proposed approach is validated and shows the practical viability. Moreover, the proposed approach can be applicable to any organizations.*

*   **Keywords:** *Rough set, almost indiscernibility, information system, fuzzy approximation space, ordering rules, information entropy*

## 1. Introduction

   A performance evaluation is a systematic and periodic process that assesses an individual organization's performance and productivity with respect to certain pre-established criteria and the objectives of the organization. It is generally applied to adhere to an organization's resources in order to accomplish maximum performance. The principal determinant of performance evaluation in an organization is its profit and loss. But, in case of educational institutions, performance evaluation deals with quality of education, assessment of students and employees, course curricula, various organizational policies, collaborative programs, industry tie-ups, infrastructure, *etc*.

   Most of the previous performance models did not take the full range of interested stakeholders into consideration and were not directly connected to strategic, quality, and financial management. A balanced scorecard approach was proposed by Cullen, Joyce *et al.*, [12] to strengthen the necessity of managing rather than just monitoring performance. Simultaneously the paper strengthened the relevance to follow private sector models for performance evaluation to deal with the most prominent quality issues. Garretson[11], in his paper, emphasized students' mindset to take into consideration so many issues apart from normal classroom activities. He reinforces the importance of the expectations of key stakeholders in the education process. Avdjieva and Wilson [14] suggested higher education institutions are required to become learning organizations where internal stakeholders also interpret and access the quality of higher education provision. Baig, Basharat, Maqsood [15] proposed an economical maturity framework which helps the existing education process to improve. Manjula *et al.*, [19] proposed a new capability maturity decision making model based on rough computing for extracting key process areas and its relevance for the development of quality education. However, they failed to study the performance evaluation of institutions in each level. Acharjya and Ezhilarasi [10] discussed the ranking of institutions and the chief factors that influence the ranking. But, the ranking should be carried out on different level of institutions. They have not considered this factor into account.

The evolution and advancement of computers and computer aided technologies lead to revolutionary changes to performance evolution systems specially in discovering huge amount of data from several target institutions and in extracting knowledge from the data by processing them. But the extracted knowledge may not always useful to its purpose, rather frequently suffers from redundancy, inconsistency *etc.*, and therefore is irrelevant. Many of the rudimentary techniques for performance evaluation are either crisp or statistical. Statistical validity is another limitation for quantifying the level of institutions. In addition to statistical validity majority of them fail with normal tests of reliability and validity. But real life situations involve with indiscernibility and both of these theories can not be applied to performance evaluation.

Fuzzy set [13] by Zadeh, rough set by Pawlak [23, 24], soft set by [7] *etc.*, are all the fruits of assiduous research works to overcome the impediments of that indiscernibility and all these are successfully applied to many fields of computer science including knowledge discovery in database, computational intelligence, knowledge engineering, granular computing *etc.*, [16, 17, 18, 20, 21] to derive knowledge in field of data analysis in which a mammoth amount of data is considered. Fuzzy set proposed by Zadeh [13] is an approach in which an element of a set belongs to the set to a degree $k$ ($0 \leq k \leq 1$), where $k$ is the membership value. The major drawback in this approach is lying in designing membership function as it requires expertise. In order to overcome this, the concept of rough set is built on the approximation of sets by a pair of sets, known as lower approximation and upper approximation which are defined in terms of equivalence relations. The lower approximation of a rough set comprises those elements of the universe which are certainly a member of the rough set. The upper approximation of a rough set consists of those elements in the universe which can possibly be classified as a member of the rough set. Further the equivalence relation is generalized to fuzzy proximity relation and rough set on fuzzy approximation space is studied by Tripathy and Acharjya [4, 8]. The introduction of fuzzy proximity relation generalizes indiscernibility to almost indiscernibility and rough set on fuzzy approximation space reduces to rough set on certain conditions. Again fuzzy proximity relation is extended to intuitionistic fuzzy proximity relation and rough set on intuitionistic fuzzy approximation space is studied by Tripathy and Acharjya [2, 9]. The different applications on these areas are explored by Tripathy and Acharjya [3, 9, 13]. In this paper, we focus on performance evaluation by using rough set on fuzzy approximation spaces, ordering rules and information entropy.

This paper uses the basic idea of rough set on fuzzy approximation space, ordering rules and information entropy to evaluate the performance of institutions. The rest of the paper is organized as follows: Section 2 presents the basics of rough set on fuzzy approximation spaces. The proposed performance evaluation approach is presented in Sections 3. In Section 4, we analyze the performance of institutions according to their levels followed by a conclusion in Section 5.

## 2. Fundamentals of Rough Set on Fuzzy Approximation Spaces

Rough set of Pawlak [23, 24] was developed as an alternative data analysis tool but subsequently emerged as a very important aspect in the areas of artificial intelligence, knowledge discovery, decision analysis, expert systems, *etc*. It can also deal with inexact, uncertain and vague datasets [18]. The basic philosophy of rough set is based on indiscernibility relation defined on a universe $U$. But indiscernibility relation based on equivalence relation is not always relevant to real life scenarios. Therefore it was necessary to make relations less stringent by excluding one or more requirements of equivalence relation and to make it more realistic. A fuzzy relation is more generalized than binary relation on a set $U$. Fuzzy proximity relation on a universal set $U$ is much general and abundant in real life situations because of exclusion of transitivity relation. So, fuzzy proximity relation has become more general, less stringent than equivalence relation and hence more practical in real life situations. Thus, the concept of fuzzy approximation space based on fuzzy proximity relation extends the efficiency and applications of rough sets on knowledge bases as discussed by Acharjya and Tripathy [4, 8].

Let $U$ be a universe. We define a fuzzy relation on $U$ as a fuzzy subset of $(U \times U)$. A fuzzy relation $R$ on $U$ is a fuzzy proximity relation if $\mu_R(x,x) = 1$ for all $x \in U$ and $\mu_R(x,y) = \mu_R(y,x)$ for $x, y \in U$. Let $R$ be a fuzzy proximity relation on $U$. Then for a given $\alpha \in [0,1]$, we say that two elements $x$ and $y$ are $\alpha$−similar with respect to $R$ if $\mu_R(x,y) \geq \alpha$ and we write $x R_\alpha y$ or $(x,y) \in R_\alpha$. Two elements $x$ and $y$ in $U$ are said to be $\alpha$−identical denoted by $x R(\alpha) y$ if either $x$ is $\alpha$−similar to $y$ or $x$ is transitively $\alpha$−similar to $y$, that is, there exists a sequence $u_1, u_2, u_3, ... , u_n$ in $U$ such that $x R_\alpha u_1, u_1 R_\alpha u_2, u_2 R_\alpha u_3, ...... , u_n R_\alpha y$. If $x$ and $y$ are $\alpha$−identical with respect to fuzzy proximity relation $R$, then we write $x R(\alpha) y$, where the relation $R(\alpha)$ for each fixed $\alpha \in [0,1]$ is an equivalence relation on $U$. The pair $(U, R)$ is called a fuzzy approximation space. The rough set of $X$, in the generated approximation space $(U, R(\alpha))$ is denoted by $(\underline{X}_\alpha, \overline{X_\alpha})$ and is defined with respect to $R_\alpha^*$, the family of equivalence classes of $R(\alpha)$. The $\alpha$−lower approximation of $X$, $\underline{X}_\alpha$ and $\alpha$−upper approximation of $X$, $\overline{X_\alpha}$ are defined as follows:

$$\underline{X}_\alpha = \cup\{Y : Y \in R_\alpha^* \text{ and } Y \subseteq X\} \tag{1}$$

$$\overline{X}_\alpha = \cup\{Y : Y \in R_\alpha^* \text{ and } Y \cap X \neq \phi\} \tag{2}$$

$X$ is said to be $\alpha$-discernible if and only if $\underline{X}_\alpha = \overline{X}_\alpha$ and $X$ is said to be $\alpha$-rough if $\underline{X}_\alpha \neq \overline{X}_\alpha$.

## 2.1. Almost Indiscernibility Relation

An information system provides all available information and knowledge about the objects under certain conditions. Objects are only perceived by using a finite number of properties without considering any semantic relationship between attribute values of a particular attribute. So trivial equality relation on values of attributes was used to do quantitative analysis as discussed in standard rough set theory [22]. But, attribute values are not usually exactly identical rather almost identical in real life situations. So Pawlak's idea of indiscernibility relation is generalized to almost indiscernibility relation which is the ground of rough set on fuzzy approximation space as discussed in the previous section.

Let $U$ be the universe and $A$ be a set of attributes. With each attribute $a \in A$ we associate a set of values $V_a$ which is called the domain of $a$. The pair $S = (U, A)$ is termed an information system. Let $B \subseteq A$. For $\alpha \in [0,1]$, we define a binary relation $R_B(\alpha)$ on $U$ defined by $x R_B(\alpha) y$ if and only if $x(a) R(\alpha) y(a)$ for all $a \in B$ where $x(a) \in V_a$ denotes the value of $x$ in $a$. It can be proved that the relation $R_B(\alpha)$ is an equivalence relation on $U$. It is also noticed that $R_B(\alpha)$ is not exact indiscernibility relation defined by Pawlak, rather it can be considered as an almost indiscernibility relation on $U$. The almost indiscernibility relation $R_B(\alpha)$ reduces to Pawlak's exact indiscernibility relation when $\alpha = 1$ and thus it generalizes Pawlak's indiscernibility relation. The family of all equivalence classes of $R_B(\alpha)$ $i.e.$, the partition generated by B for $\alpha \in [0,1]$ is denoted by $U / R_B(\alpha)$. If $(x, y) \in R_B(\alpha)$, then we say that $x$ and $y$ are $\alpha -$ indiscernible. These are the basic building blocks of rough set on fuzzy approximation space.

## 2.2. Ordered Information System

The basis of data mining is to acquire knowledge by classifying objects. In the previous section, we have mentioned about an information system and the almost indiscernibility relation that is used in classification. But, in real life situations, we may face many problems that are not simply classification. One such type of problems is the ordering of objects and its attribute values under certain consideration. An information system is defined as a quadruple $I = (U, A, V_a, f_a)$ where $U$ is a finite nonempty set of objects called the universe, $A$ is a finite nonempty set of attributes, $V_a$ is a nonempty set of values for $a \in A$, $f_a : U \to V_a$ is an information function.

Consider the sample information system given in Table 1. Here, we have A = {Configure price ($cp$), Weight ($w$), Battery life ($bl$), Battery replace ability ($br$), HDD speed ($hs$), Optical drive ($od$)} and $V_{Cp}$ = {$1800, $1500, $2000, $1860, $2100}. Likewise we have $V_W$ = {3lbs., 3.97lbs., 2.4lbs., 4lbs., 2.7lbs.}, $V_{Bl}$ = {5 hours, 3 hours, 8 hours, 2.5 hours}, $V_{Br}$ = {Yes, No}, $V_{Hs}$ = {4200 rpm, 5400 rpm}, and $V_{Od}$ = {Yes, No}.

**Table 1. Sample Information System**

| Object | Cp (Dollars) | W (lbs) | Bl (Hours) | Br | Hs (Rpm) | Od |
|---|---|---|---|---|---|---|
| Apple Mac Book ($x_1$) | $1800 | 3 | 5 | No | 4200 | No |
| Dell XPS ($x_2$) | $1500 | 3.97 | 3 | Yes | 5400 | Yes |
| Toshiba Portege ($x_3$) | $2000 | 2.4 | 8 | Yes | 5400 | Yes |
| Fujitsu Life Book ($x_4$) | $1860 | 4 | 2.5 | Yes | 5400 | Yes |
| Sony Vaio ($x_5$) | $2100 | 2.7 | 5 | Yes | 5400 | Yes |

An ordered information system is defined as $OIS = \{I, \{\prec_a : a \in A\}\}$ where, I is a standard information system and $\prec_a$ is an order relation on attribute $a \in A$. An ordering of values of a particular attribute a naturally induces an ordering of objects:

$$x \prec_{\{a\}} y \iff f_a(x) \prec_a f_a(y) \tag{3}$$

where, $\prec_{\{a\}}$ denotes an order relation on $U$ induced by the attribute $a$. An object $x_i$ is ranked ahead of object $x_j$ if and only if the value of $x_i$ on the attribute $a$ is ranked ahead of the value of $x_j$ on the attribute $a$. For example, information system given above in Table 1 becomes order information system on introduction of the following ordering relations:

$\prec_{Cp}$:    $2100 \prec 2000 \prec 1860 \prec 1800 \prec 1500$

$\prec_{W}$:    $4 \prec 3.97 \prec 3 \prec 2.7 \prec 2.4$

$\prec_{Bi}$:    $8 \prec 5 \prec 3 \prec 2.5$

$\prec_{Br}$:    $\text{Yes} \prec \text{No}$

$\prec_{Hs}$:    $5400 \prec 4200$

$\prec_{Od}$:    $\text{Yes} \prec \text{No}$

For a subset of attributes $B \subseteq A$, an object $x_i$ is ranked ahead of $x_j$ if and only if $x_i$ is ranked ahead of $x_j$ according to all attributes in $B$, *i.e.*,

$$x \prec_B y \iff f_a(x) \prec_a f_a(y) \quad \forall a \in B$$

$$\iff \wedge_{a \in B} f_a(x) \prec_a f_a(y) \iff \cap_{a \in B} \{\prec_a\}$$

The above definition is a straightforward generalization of the standard definition of equivalence relations in rough set theory, where the equality relation is used.

## 3. Proposed Evaluation Index System

In this section, we propose evaluation index system that can be used to access the performance evaluation of all the educational institutions in the universe for any level proposed in the model. There are many maturity hierarchical models developed and analyzed by different researchers [1, 5, 15]. Some of these models are used in obtaining rank of the institutions. But, they are lacking in obtaining performance evaluation of institutions. Many authors proposed a list of quantitative and qualitative matrices that can be used to access the rank of any institutions, such as intellectual capital, infrastructure facility, placement performance, student satisfaction, international linkage *etc*. The proposed evaluation index system, as shown in Figure 1, is divided into three layers such as objective layer, standard layer and metric layer. The standard layer contains four primary evaluation indexes such as level 1, level 2, level 3 and level 4. Literature and numerical values based on different levels were collected and studied. These parameters forms the metric set for our analysis. The metric set consisting of these secondary evaluation indexes form the metric layer. The metric set that plays a vital role in performance evaluation and the notations that are used in our analysis are given in Table 2.

**Table 2. Notation Representation Table**

| Level | Metric Set | Abbreviation | Notation |
|---|---|---|---|
| Level 1 | Intellectual Capital | IC | $a_1$ |
| | Infrastructure Facility | IF | $a_2$ |
| | Placement Performance | PP | $a_3$ |
| | Fee (In Lakh Rupees) | Fee | $a_4$ |
| | Course Curriculum | CC | $a_5$ |
| Level 2 | Technical and Financial Aids | TFA | $a_6$ |
| | Average Salary Offered | ASO | $a_7$ |
| | Student Satisfaction | SS | $a_8$ |
| | Teaching Learning Assessment | TLA | $a_9$ |
| | Extra Curricular Activities | ECA | $a_{10}$ |
| Level 3 | International Linkage | IL | $a_{11}$ |
| | Recruiters Satisfaction | RS | $a_{12}$ |
| | Industry Institute Interface | III | $a_{13}$ |
| Level 4 | Research, Consultancy and Extension | RCE | $a_{14}$ |
| | Teaching and Learning Practices | TLP | $a_{15}$ |
| | Maturity and Stability | MS | $a_{16}$ |

In Table 1 of this paper, we have discussed an information system. The information system contains data about the universe $U$, and the attributes (metric). Our objective is to measure the performance of educational institutions that will help us in identifying good institutions in each level. Further analysis can be done as discussed by the author [19] to obtain the rank of the institutions for each level. We normalize the quantitative data by using rough set on fuzzy approximation space with ordering and formed a classifying rule to categorize the metric set values into different groups.

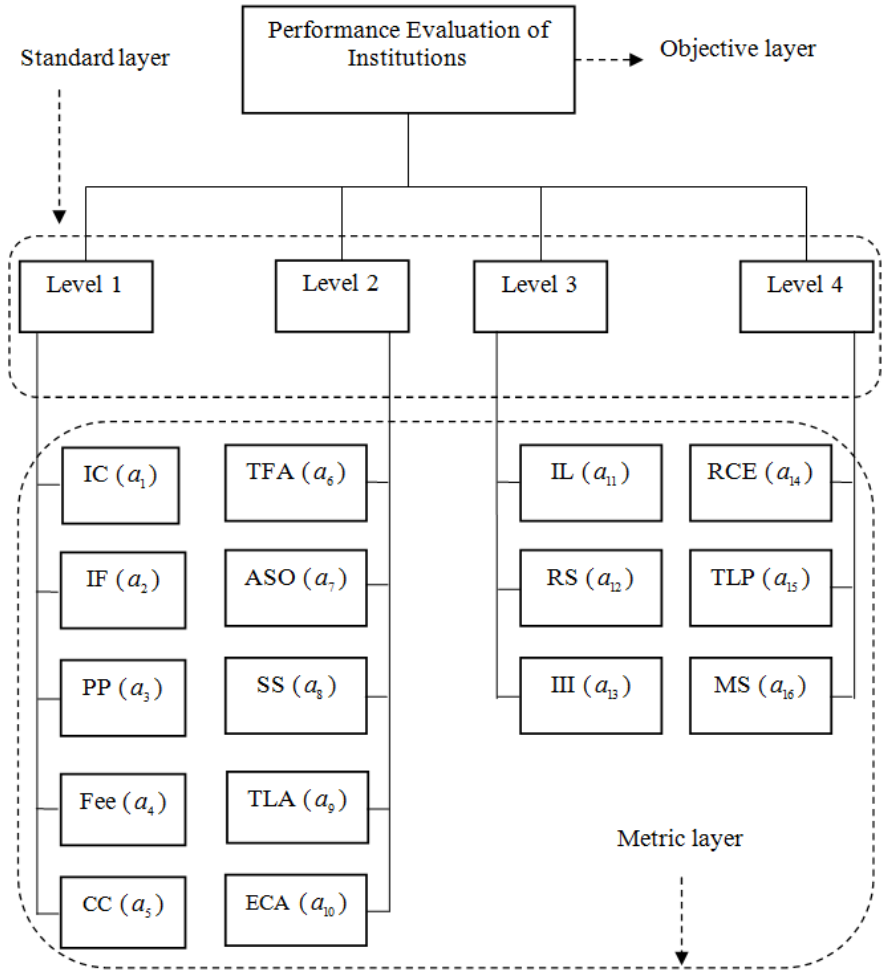

**Figure 1. Proposed Evaluation Index System**

## 3.1. Determination of Weighting Coefficients

Let $U$ be a universe, and $A = \{a_1, a_2, a_3, \cdots, a_m\}$ be the attribute set. Let $B \subseteq A$. Let $R_B$ be a fuzzy proximity relation defined over $U$. For $\alpha \in [0,1]$, the family of almost equivalence classes $R_B(\alpha)$, *i.e.*, the partition generated by $B$ for $\alpha \in [0,1]$ can be expressed as:

$$U / R_B(\alpha) = \{X_1, X_2, X_3, \cdots, X_n\}$$

Therefore, the probability distribution constituted by $B \subseteq A$ is given as follows:

$$[U / R_B(\alpha); p] = \begin{bmatrix} X_1 & X_2 & X_3 & \cdots & X_n \\ p(X_1) & p(X_2) & p(X_3) & \cdots & p(X_n) \end{bmatrix}$$

where $p$ is the probability defined as $p(X_i) = \dfrac{|X_i|}{|U|}$ for $i = 1, 2, 3, \cdots, n$. Thus the information entropy [6] of the attribute set $B$, $H(B)$, is defined as:

$$H(B) = -\sum_{i=1}^{n} p(X_i) \log p(X_i) \qquad (4)$$

Let $OIS = \{U, A, V_a, f_a, \{\prec_a : a \in A\}\}$ be an ordered information system. The significance of an attribute $a_i \in A$ can be defined as:

$$SGF_A(a_i) = |H(A) - H(A - \{a_i\})| \qquad (5)$$

An attribute $a_i \in A$ is said to be non redundant if $SGF_A(a_i) > 0$. If $SGF_A(a_i) = 0$, then $a$ is called a reduct of $A$ and removal of $a_i$ in $A$ does not affect the performance evaluation. Therefore, the weighting coefficient of performance evaluation index ($w$) is given as:

$$w_{a_k} = \frac{SGF_A(a_k)}{\sum_{k=1}^{m} SGF_A(a_k)}, \ k = 1, 2, 3, \cdots, m \qquad (6)$$

### 3.2. Performance Evaluation Algorithm

In this section we propose a performance evaluation algorithm that generates the performance of all institutions in each level by eliminating all dispensable secondary evaluation indexes of the metric layer from the information system. It also identifies the institutions having better performance in each level. We apply the following steps in order to generate the performance of institutions.

**Algorithm**

Input:      Information system

Output:    Performance Evaluation

1. Construct an evaluation index system as shown in Figure 1.
2. Obtain the data classification for each secondary evaluation index of the metric layer by using fuzzy proximity relation.
3. Determine the comment set $\phi = \{\phi_1, \phi_2, \phi_3, \phi_4, \cdots, \phi_s\}$ for each secondary evaluation index and the corresponding grades are set to be $\{s, s-1, s-2, \cdots, 1\}$.
4. Impose ordering rules to the classification and construct the order information system.
5. Suppose that there are $n$ ($n \geq 1$) institutions to be evaluated, the universe can be denoted by $U = \{x_1, x_2, x_3, \cdots, x_n\}$. Obtain the partition $U / R_A$ caused by the secondary evaluation index. Also, compute $U / R_{A-\{a_i\}}$ for $i = 1, 2, 3, \cdots, m$.
6. Compute the information entropy $H(A)$ and $H(A - \{a_i\})$ of the attribute set $A$ and $(A - \{a_i\})$ respectively by using equation 4.
7. Determine the significance of each attribute $a_i \in A$. Obtain the non redundant attribute. For each non redundant attribute, compute the weighting coefficient of performance evaluation index ($w$) by using equation 6.
8. The performance evaluation value of r[th] institution ($W_r$) can be formulated as:

$$W_r = \sum w_{a_k} \cdot V_{a_k}^r \ \text{ for } k = 1, 2, 3, \cdots, m \qquad (7)$$

## 4. An Empirical Study on Performance Evaluation

In this section, we demonstrate how the above concept can be applied to real life situation. We consider the example in which we evaluate the performance of an institution in a group of institutions. In the Table 3 given below, we consider few secondary evaluation indexes to evaluate the performance of an institution; their possible range of values and a fuzzy proximity relation which characterized the relationship between secondary evaluation indexes. The institution with high score in each secondary evaluation indexes becomes an ideal case for better performance. But, such type of cases is rare in practice. So, an institution may not excel in all the secondary evaluation indexes in order to get top position in each level. However, out of these secondary evaluation indexes, some may have greater influence on the evaluation than the others. For different values of membership function $\alpha$, these secondary evaluation indexes may be different. In fact, if we decrease the value of $\alpha$, more and more number of secondary evaluation indexes shall become indispensable. The membership functions have been designed such that their values should lie in [0, 1] and also these functions must be symmetric.

Now, we define a fuzzy proximity relation $R(i_j, i_k)$ in order to identify the almost indiscernibility among the objects $i_j$ and $i_k$, where

$$R(i_j, i_k) = 1 - \frac{|V_{i_j} - V_{i_k}|}{\text{Range}} \qquad (8)$$

The institutes can be judged by the outputs which are produced. The quality of the output can be judged by the placement performance of the institute and is given the highest weight with a score 385 which is around 21% of total weight. To produce the quality output the input should be of high quality. The major inputs for an institute are intellectual capital and infrastructure facilities for providing quality education. Accordingly the scores for intellectual capital and infrastructure facility are fixed as 250 and 200 respectively that are 12% and 11% of total weight. The study expense in general is from 1 Lakh to 6 Lakhs in Indian Rupee. The student placed in the company is expected to serve the company up to their expectation, *i.e.*, the student shall meet the recruiter's satisfaction which is given a score of 180 which is around 9% of the total weight. At the same time student's satisfaction and extra curricular activities, which play a vital role for prospective students, are given scores 60 and 80 respectively of weights 4% and 6% of the total weight. Technical and financial aids, which is essential at the growth level, is given a score 70 which is around 5% of total weight. Similarly, to check the performance of institutions at higher level, the secondary evaluation indexes such as international linkage, industry institute interface, research consultancy and extension, and maturity and stability have been given scores 200, 90, 200 and 60 respectively of weights 10%, 8%, 10% and 4% of the total weight. However, out of these secondary evaluation indexes, some indexes may have greater influence than the rest. But, the index values of these secondary evaluation indexes obtained are almost indiscernible and hence can be classified by using rough set on fuzzy approximation space [4] and ordering rules. In view of the length of the paper and to make our analysis simple, we consider a small universe of 10 institutions and the information pertaining to them are presented in the Table 4. We keep the identities of the institutions confidential as they do not affect our analysis. The data collected is considered to be the representative figure and tabulated below in Table 4.

**Table 3. Notation Representation Table**

| Level | Abbreviation | Notation | Possible weight | Range |
|---|---|---|---|---|
| Level 1 | IC | $a_1$ | [0 – 250] | 250 |
| | IF | $a_2$ | [0 – 200] | 200 |
| | PP | $a_3$ | [0 – 385] | 385 |
| | Fee | $a_4$ | [0 – 6] | 6 |
| | CC | $a_5$ | Very good, Good, Average | -- |
| Level 2 | TFA | $a_6$ | [0 – 70] | 70 |
| | ASO | $a_7$ | [0 – 7] | 7 |
| | SS | $a_8$ | [0 – 60] | 60 |
| | ECA | $a_9$ | [0 – 80] | 80 |
| | TLA | $a_{10}$ | Very good, Good, Average | -- |
| Level 3 | IL | $a_{11}$ | [0 – 200] | 200 |
| | RS | $a_{12}$ | [0 – 180] | 180 |
| | III | $a_{13}$ | [0 – 90] | 90 |
| Level 4 | RCE | $a_{14}$ | [0 – 200] | 200 |
| | MS | $a_{15}$ | [0 – 60] | 60 |
| | TLP | $a_{16}$ | Very good, Good, Average | -- |

**Table 4. Small Universe of Institutions**

| Institution | IC | IF | PP | Fee | CC | TFA | ASO | SS | ECA | TLA | IL | RS | III | RCE | MS | TLP |
|---|---|---|---|---|---|---|---|---|---|---|---|---|---|---|---|---|
| $i_1$ | 229 | 151 | 303 | 5.6 | Very good | 69 | 6.2 | 56 | 69.37 | Very good | 197 | 175 | 88 | 158 | 35 | Average |
| $i_2$ | 130 | 120 | 209 | 4.8 | Very good | 17 | 3 | 54 | 26.03 | Good | 124 | 125 | 61 | 113 | 15 | Average |
| $i_3$ | 226 | 145 | 266 | 5.4 | Very good | 67 | 6.25 | 53 | 78.86 | Very good | 114 | 120 | 62 | 125 | 30 | Good |
| $i_4$ | 100 | 113 | 151 | 3.9 | Good | 37 | 3.4 | 53 | 29.52 | Average | 120 | 96 | 37 | 122 | 7 | Average |
| $i_5$ | 97 | 115 | 245 | 4.7 | Good | 20 | 2.8 | 51 | 25.03 | Average | 161 | 105 | 59 | 160 | 28 | Good |
| $i_6$ | 227 | 142 | 298 | 5.3 | Very good | 62 | 6 | 48 | 67.86 | Very good | 196 | 176 | 85 | 195 | 57 | Very good |
| $i_7$ | 96 | 75 | 144 | 4.0 | Average | 36 | 2.1 | 39 | 10.05 | Average | 159 | 121 | 63 | 163 | 41 | Average |
| $i_8$ | 211 | 117 | 247 | 5.3 | Very good | 51 | 4.9 | 47 | 48.57 | Good | 105 | 80 | 24 | 157 | 31 | Good |
| $i_9$ | 110 | 59 | 138 | 4.2 | Average | 35 | 3.7 | 40 | 30.12 | Average | 163 | 77 | 55 | 105 | 18 | Average |
| $i_{10}$ | 217 | 135 | 251 | 5.1 | Very good | 50 | 4.8 | 42 | 44.56 | Good | 100 | 75 | 40 | 90 | 10 | Average |

## 4.1. Results and Analysis

In this section, we discuss in detail the subsequent steps of the performance evaluation algorithm for the empirical study taken under consideration. A target dataset for analysis as shown in Table 4 is considered. We have designed fuzzy proximity relations based on the secondary evaluation indexes and computed the almost similarity between them. The fuzzy proximity relation identifies the almost indiscernibility among the institutions. This result induces the equivalence classes. We obtain categorical classes on imposing order relation on this classification. Considering the length of the paper, we compute the performance of institutions according to the secondary evaluation indexes at level 1. The fuzzy proximity relations $R_i, i = 1, 2, 3, 4$ corresponding to the attributes IC, IF, PP and Fee are given in tables 5, 6, 7 and 8 respectively. Let $R_5$ be the relation corresponding to the attribute CC.

**Table 5. Fuzzy Proximity Relation for IC**

| $R_1$ | $i_1$ | $i_2$ | $i_3$ | $i_4$ | $i_5$ | $i_6$ | $i_7$ | $i_8$ | $i_9$ | $i_{10}$ |
|---|---|---|---|---|---|---|---|---|---|---|
| $i_1$ | 1.000 | .604 | .988 | 484 | .472 | .992 | .468 | .928 | .524 | .952 |
| $i_2$ | .604 | 1.000 | .616 | .880 | .868 | .612 | .864 | .676 | .920 | .652 |
| $i_3$ | .988 | .616 | 1.000 | .496 | .484 | .996 | .480 | .940 | .536 | .964 |
| $i_4$ | 484 | .880 | .496 | 1.000 | .988 | .492 | .984 | .556 | .960 | .532 |
| $i_5$ | .472 | .868 | .484 | .988 | 1.000 | .480 | .996 | .544 | .948 | .520 |
| $i_6$ | .992 | .612 | .996 | .492 | .480 | 1.000 | .476 | .936 | .532 | .960 |
| $i_7$ | .468 | .864 | .480 | .984 | .996 | .476 | 1.000 | .540 | .944 | .516 |
| $i_8$ | .928 | .676 | .940 | .556 | .544 | .936 | .540 | 1.000 | .596 | .976 |
| $i_9$ | .524 | .920 | .536 | .960 | .948 | .532 | .944 | .596 | 1.000 | .572 |
| $i_{10}$ | .952 | .652 | .964 | .532 | .520 | .960 | .516 | .976 | .572 | 1.000 |

**Table 6. Fuzzy Proximity Relation for IF**

| $R_2$ | $i_1$ | $i_2$ | $i_3$ | $i_4$ | $i_5$ | $i_6$ | $i_7$ | $i_8$ | $i_9$ | $i_{10}$ |
|---|---|---|---|---|---|---|---|---|---|---|
| $i_1$ | 1.000 | .845 | .970 | .810 | .820 | .955 | .620 | .830 | .540 | .920 |
| $i_2$ | .845 | 1.000 | .875 | .965 | .975 | .890 | .775 | .985 | .695 | .925 |
| $i_3$ | .970 | .875 | 1.000 | .840 | .850 | .985 | .650 | .860 | .570 | .950 |
| $i_4$ | .810 | .965 | .840 | 1.000 | .990 | .855 | .810 | .980 | .730 | .890 |
| $i_5$ | .820 | .975 | .850 | .990 | 1.000 | .865 | .800 | .990 | .720 | .900 |
| $i_6$ | .955 | .890 | .985 | .855 | .865 | 1.000 | .665 | .875 | .585 | .965 |
| $i_7$ | .620 | .775 | .650 | .810 | .800 | .665 | 1.000 | .790 | .920 | .700 |
| $i_8$ | .830 | .985 | .860 | .980 | .990 | .875 | .790 | 1.000 | .710 | .910 |
| $i_9$ | .540 | .695 | .570 | .730 | .720 | .585 | .920 | .710 | 1.000 | .620 |
| $i_{10}$ | .920 | .925 | .950 | .890 | .900 | .965 | .700 | .910 | .620 | 1.000 |

## Table 7. Fuzzy Proximity Relation for PP

| $R_3$ | $i_1$ | $i_2$ | $i_3$ | $i_4$ | $i_5$ | $i_6$ | $i_7$ | $i_8$ | $i_9$ | $i_{10}$ |
|---|---|---|---|---|---|---|---|---|---|---|
| $i_1$ | 1.000 | .756 | .904 | .605 | .849 | .987 | .587 | .855 | .571 | .865 |
| $i_2$ | .756 | 1.000 | .852 | .849 | .906 | .769 | .831 | .901 | .816 | .891 |
| $i_3$ | .904 | .852 | 1.000 | .701 | .945 | .917 | .683 | .951 | .668 | .961 |
| $i_4$ | .605 | .849 | .701 | 1.000 | .756 | .618 | .982 | .751 | .966 | .740 |
| $i_5$ | .849 | .906 | .945 | .756 | 1.000 | .862 | .738 | .995 | .722 | .984 |
| $i_6$ | .987 | .769 | .917 | .618 | .862 | 1.000 | .600 | .868 | .584 | .878 |
| $i_7$ | .587 | .831 | .683 | .982 | .738 | .600 | 1.000 | .732 | .984 | .722 |
| $i_8$ | .855 | .901 | .951 | .751 | .995 | .868 | .732 | 1.000 | .717 | .990 |
| $i_9$ | .571 | .816 | .668 | .966 | .722 | .584 | .984 | .717 | 1.000 | .706 |
| $i_{10}$ | .865 | .891 | .961 | .740 | .984 | .878 | .722 | .990 | .706 | 1.000 |

## Table 8. Fuzzy Proximity Relation for Fee

| $R_4$ | $i_1$ | $i_2$ | $i_3$ | $i_4$ | $i_5$ | $i_6$ | $i_7$ | $i_8$ | $i_9$ | $i_{10}$ |
|---|---|---|---|---|---|---|---|---|---|---|
| $i_1$ | 1.000 | .867 | .967 | .717 | .850 | .950 | .733 | .950 | .767 | .917 |
| $i_2$ | .867 | 1.000 | .900 | .850 | .983 | .917 | .867 | .917 | .900 | .950 |
| $i_3$ | .967 | .900 | 1.000 | .750 | .883 | .983 | .767 | .983 | .800 | .950 |
| $i_4$ | .717 | .850 | .750 | 1.000 | .867 | .767 | .983 | .767 | .950 | .800 |
| $i_5$ | .850 | .983 | .883 | .867 | 1.000 | .900 | .883 | .900 | .917 | .933 |
| $i_6$ | .950 | .917 | .983 | .767 | .900 | 1.000 | .783 | 1.000 | .817 | .967 |
| $i_7$ | .733 | .867 | .767 | .983 | .883 | .783 | 1.000 | .783 | .967 | .817 |
| $i_8$ | .950 | .917 | .983 | .767 | .900 | 1.000 | . 783 | 1.000 | .817 | .967 |
| $i_9$ | .767 | .900 | .800 | .950 | .917 | .817 | .967 | .817 | 1.000 | .850 |
| $i_{10}$ | .917 | .950 | .950 | .800 | .933 | .967 | .817 | .967 | .850 | 1.000 |

Now on considering the almost similarity of 85% *i.e.*, $\alpha \geq 0.85$ it is observed from Table 5 that $R_1(i_1, i_1) = 1$ ; $R_1(i_1, i_3) = 0.988$ ; $R_1(i_1, i_6) = 0.992$ ; $R_1(i_1, i_{10}) = 0.952$ ; $R_1(i_2, i_2) = 1$ ; $R_1(i_2, i_4) = 0.88$ ; $R_1(i_2, i_5) = 0.868$ ; $R_1(i_2, i_7) = 0.864$ ; and $R_1(i_2, i_9) = 0.920$ . Thus, the institutions $i_1, i_3, i_6, i_8, i_{10}$ are $\alpha$-identical. Similarly, $i_2, i_4, i_5, i_7, i_9$ are $\alpha$-identical. Thus, we get

$$U/R_1^\alpha = \{\{i_1, i_3, i_6, i_8, i_{10}\}, \{i_2, i_4, i_5, i_7, i_9\}\}$$

Therefore, the values of the secondary evaluation index IC are classified into two categories namely very good and good and hence can be ordered. Similarly, the different equivalence classes obtained from Table 6, 7, 8 corresponding to the attributes IF, PP, and Fee are given below.

$$U/R_2^\alpha = \{\{i_1, i_2, i_3, i_4, i_5, i_6, i_8, i_{10}\}, \{i_7, i_9\}\}$$

$$U/R_3^\alpha = \{\{i_1, i_2, i_3, i_5, i_6, i_8, i_{10}\}, \{i_4, i_7, i_9\}\}$$

$$U/R_4^\alpha = \{i_1, i_2, i_3, i_4, i_5, i_6, i_7, i_8, i_9, i_{10}\}$$

$$U/R_5^\alpha = \{\{i_1, i_2, i_3, i_6, i_8, i_{10}\}, \{i_4, i_5\}, \{i_7, i_9\}\}$$

From the above classification, it is clear that the values of the secondary evaluation indexes IF and PP are classified into two categories namely very good and good. The values of the secondary evaluation index fee are classified into a single category such as very good. Similarly, the values of secondary evaluation index CC are classified into three categories namely very good, good and average. Therefore according to step 3, we use a comment set $\phi$ = {Very good, Good, Average} and the corresponding grades are set to be {3, 2, 1}. Thus, the ordered information system for level 1 of the small universe of institutions is given below in Table 9.

**Table 9. Order Information System for Level 1**

| Institutes | IC | IF | PP | Fee | CC |
|---|---|---|---|---|---|
| $i_1$ | Very good (3) | Very good (3) | Very good (3) | Good (2) | Very good (3) |
| $i_2$ | Good (2) | Very good (3) | Very good (3) | Good (2) | Very good (3) |
| $i_3$ | Very good (3) | Very good (3) | Very good (3) | Good (2) | Very good (3) |
| $i_4$ | Good (2) | Very good (3) | Good (2) | Good (2) | Good (2) |
| $i_5$ | Good (2) | Very good (3) | Very good (3) | Good (2) | Good (2) |
| $i_6$ | Very good (3) | Very good (3) | Very good (3) | Good (2) | Very good (3) |
| $i_7$ | Good (2) | Good (2) | Good (2) | Good (2) | Average (1) |
| $i_8$ | Very good (3) | Very good (3) | Very good (3) | Good (2) | Very good (3) |
| $i_9$ | Good (2) | Good (2) | Good (2) | Good (2) | Average (1) |
| $i_{10}$ | Very good (3) | Very good (3) | Very good (3) | Good (2) | Very good (3) |

$\prec_{IC}$: Very Good $\prec$ Good

$\prec_{IF}$: Very Good $\prec$ Good

$\prec_{PP}$: Very Good $\prec$ Good

$\prec_{Fee}$: Good

$\prec_{CC}$: Very Good $\prec$ Good $\prec$ Average

Let us consider A = {IC, IF, PP, Fee, CC}. Let $R$ be the equivalence relation defined on A. Therefore, the partition caused by the secondary evaluation index A for level 1 is given below.

$$U / R_A = \{\{i_1, i_3, i_6, i_8, i_{10}\}, \{i_2\}, \{i_4\}, \{i_5\}, \{i_7, i_9\}\}$$

Similarly, we get

$$U/R_{A-\{a_1\}} = \{\{i_1, i_2, i_3, i_6, i_8, i_{10}\}, \{i_4\}, \{i_5\}, \{i_7, i_9\}\}$$

$$U/R_{A-\{a_2\}} = \{\{i_1, i_3, i_6, i_8, i_{10}\}, \{i_2\}, \{i_4\}, \{i_5\}, \{i_7, i_9\}\}$$

$$U/R_{A-\{a_3\}} = \{\{i_1, i_3, i_6, i_8, i_{10}\}, \{i_2\}, \{i_4, i_5\}, \{i_7, i_9\}\}$$

$$U/R_{A-\{a_4\}} = \{\{i_1, i_3, i_6, i_8, i_{10}\}, \{i_2\}, \{i_4\}, \{i_5\}, \{i_7, i_9\}\}$$

$$U/R_{A-\{a_5\}} = \{\{i_1, i_3, i_6, i_8, i_{10}\}, \{i_2, i_5\}, \{i_4\}, \{i_7, i_9\}\}$$

The information entropy $H(A)$ and $H(A-\{a_i\})$ of the attribute set $A$ and $(A-\{a_i\})$ are calculated and given as: $H(A) = 0.59$, $H(A-\{a_1\}) = 0.473$, $H(A-\{a_2\}) = 0.59$, $H(A-\{a_3\}) = 0.53$, $H(A-\{a_4\}) = 0.59$ and $H(A-\{a_5\}) = 0.53$. The significance of each attribute $a_i \in A$ is computed and given as: $SGF_A(a_1) = 0.117$, $SGF_A(a_2) = 0 = SGF_A(a_4)$, $SGF_A(a_3) = 0.06$ and $SGF_A(a_5) = 0.06$. From the above analysis, it is clear that the attributes $a_2$ and $a_4$ are redundant and the weighting coefficient of performance evaluation index of other attributes are calculated as: $w_1 = 0.494$, $w_3 = 0.253$ and $w_5 = 0.253$. So we obtain the performance evaluation values ($W_r$) for level 1 of the ten institutions which are shown in Table 10.

**Table 10. The Performance Evaluation Values for Level 1**

| Institution | $i_1$ | $i_2$ | $i_3$ | $i_4$ | $i_5$ | $i_6$ | $i_7$ | $i_8$ | $i_9$ | $i_{10}$ |
|---|---|---|---|---|---|---|---|---|---|---|
| Value | 3 | 2.012 | 3 | 1.253 | 1.759 | 3 | 1 | 3 | 1 | 3 |

Similarly for level 2, on considering the attributes A = {TFA, ASO, SS, ECA, TLA}, the information entropy $H(A)$ and $H(A-\{a_i\})$ of the attribute set $A$ and $(A-\{a_i\})$ are calculated and given as: $H(A) = 0.736$, $H(A-\{a_6\}) = 0.654$, $H(A-\{a_7\}) = 0.736$, $H(A-\{a_8\}) = 0.736$, $H(A-\{a_9\}) = 0.676$ and $H(A-\{a_{10}\}) = 0.654$. The significance of each attribute $a_i \in A$ is computed and given as: $SGF_A(a_6) = 0.082$, $SGF_A(a_7) = 0 = SGF_A(a_8)$, $SGF_A(a_9) = 0.06$ and $SGF_A(a_{10}) = 0.082$. From the above analysis, it is clear that the attributes $a_7$ and $a_8$ are redundant and the weighting coefficient of performance evaluation index of other attributes are calculated as: $w_{a_6} = 0.366$, $w_{a_9} = 0.268$ and $w_{a_{10}} = 0.366$. So we obtain the performance evaluation values ($W_r$) for level 2 of the ten institutions which are shown in Table 11.

**Table 11. The Performance Evaluation Values for Level 2**

| Institution | $i_1$ | $i_2$ | $i_3$ | $i_4$ | $i_5$ | $i_6$ | $i_7$ | $i_8$ | $i_9$ | $i_{10}$ |
|---|---|---|---|---|---|---|---|---|---|---|
| Value | 4 | 1.902 | 4 | 2 | 1.634 | 4 | 1.634 | 3 | 2 | 3 |

Similarly for level 3, on considering the attributes A = {IL, RS, III}, the information entropy $H(A)$ and $H(A-\{a_i\})$ of the attribute set $A$ and $(A-\{a_i\})$ are calculated and given as: $H(A) = 0.594$, $H(A - \{a_{11}\}) = 0.448$, $H(A - \{a_{12}\}) = 0.594$ and $H(A - \{a_{13}\}) = 0.448$. The significance of each attribute $a_i \in A$ is computed and given as: $SGF_A(a_{11}) = 0.146$, $SGF_A(a_{12}) = 0$ and $SGF_A(a_{13}) = 0.146$. From the above analysis, it is clear that the attribute $a_{12}$ is redundant and the weighting coefficient of performance evaluation index of other attributes are calculated as: $w_{a_{11}} = 0.5$ and $w_{a_{13}} = 0.5$. So we obtain the performance evaluation values ($W_r$) for level 3 of the ten institutions which are shown in Table 12.

### Table 12. The Performance Evaluation Values for Level 3

| Institution | $i_1$ | $i_2$ | $i_3$ | $i_4$ | $i_5$ | $i_6$ | $i_7$ | $i_8$ | $i_9$ | $i_{10}$ |
|---|---|---|---|---|---|---|---|---|---|---|
| Value | 3 | 1.5 | 1.5 | 1 | 2 | 3 | 2 | 1 | 2 | 1 |

Similarly for level 4, on considering the attributes A = {RCE, MS, TLP}, the information entropy $H(A)$ and $H(A - \{a_i\})$ of the attribute set $A$ and $(A - \{a_i\})$ are calculated and given as: $H(A) = 0.64$, $H(A - \{a_{14}\}) = 0.557$, $H(A - \{a_{15}\}) = 0.64$ and $H(A - \{a_{16}\}) = 0.518$. The significance of each attribute $a_i \in A$ is computed and given as: $SGF_A(a_{14}) = 0.083$, $SGF_A(a_{15}) = 0$ and $SGF_A(a_{16}) = 0.122$. From the above analysis, it is clear that the attribute $a_{15}$ is redundant and the weighting coefficient of performance evaluation index of other attributes are calculated as: $w_{a_{14}} = 0.405$ and $w_{a_{16}} = 0.595$. So we obtain the performance evaluation values ($W_r$) for level 4 of the ten institutions which are shown in Table 13.

### Table 13. The Performance Evaluation Values for Level 4

| Institution | $i_1$ | $i_2$ | $i_3$ | $i_4$ | $i_5$ | $i_6$ | $i_7$ | $i_8$ | $i_9$ | $i_{10}$ |
|---|---|---|---|---|---|---|---|---|---|---|
| Value | 1.405 | 1 | 1.595 | 1 | 2 | 3 | 1.405 | 2 | 1 | 1 |

Figure 2 illustrates the overall performance evaluation result of the ten institutions according to the evaluation index system shown in Figure 1. According to level 1, institutions $i_1, i_3, i_6, i_8$ and $i_{10}$ outperform others, and institutions $i_7$ and $i_9$ are inefficient in overall performance in terms of level 1. Secondly, institutions $i_1, i_3$ and $i_6$ outperform others, and institutions $i_5$ and $i_7$ are inefficient in overall performance in terms of level 2. Thirdly, in level 3, institutions $i_1$ and $i_6$ outperform others, and institutions $i_4, i_8$ and $i_{10}$ are inefficient in overall performance in terms of level 3. Finally, in level 4, institution $i_6$ has better performance than others due to the highest research, consultancy and extension (RCE).

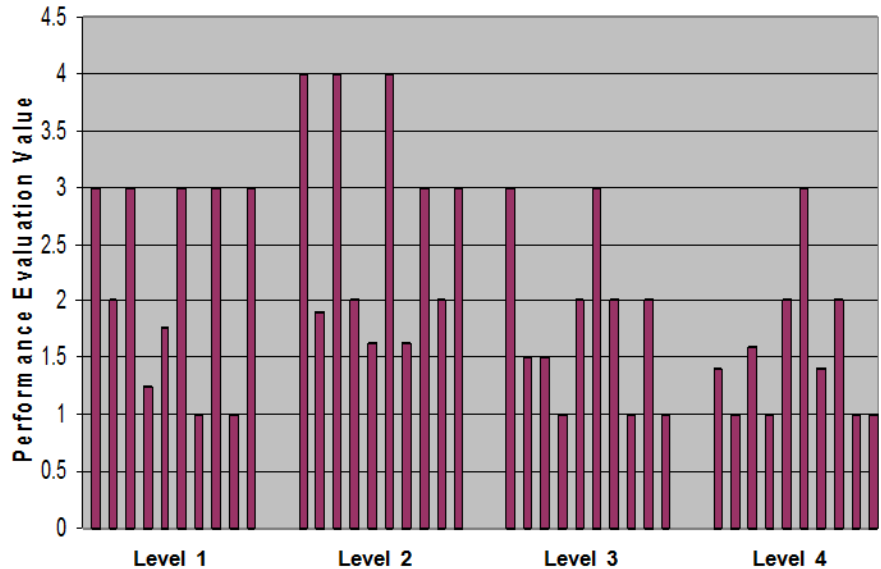

**Figure 2. Overall Performance Evaluations of Ten Institutions**

## 5. Conclusion and Future Extension

In this paper, we consider the issues of performance evaluation approach for educational institutions in various levels. We have also focused on effectiveness and fairness in the design of evaluation approach. We proposed a rough computing based performance evaluation approach for educational institutions. This approach consists of a three layer evaluation index system and uses rough set on fuzzy approximation space with ordering, information entropy to explore the evaluation index data. We achieve the performance evaluation for educational institutions through the proposed approach. An empirical study is taken into consideration to illustrate the effectiveness and feasibility of the proposed research. In addition, our future work aims at attaining chief secondary evaluation indexes affecting the decisions at each level. We also aim to establish a modularized and user friendly computer evaluation system to access the performance of educational institutions.

# Authors

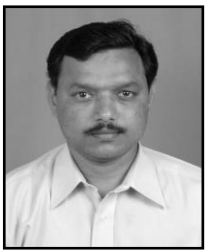

**D. P. Acharjya** received his Ph. D in computer science from Berhampur University; M. Tech. degree in computer science from Utkal University, India; M. Phil. from Berhampur University, India; and M. Sc. from NIT, Rourkela, India. He has been awarded with Gold Medal in M. Sc. At present, he is working as a Professor in the school of computing sciences and engineering, VIT University, Vellore, India. He has authored many national and international journal papers and four books; Fundamental Approach to Discrete Mathematics, Computer Based on Mathematics, Theory of Computation; Rough Set in Knowledge Representation and Granular Computing to his credit. He is associated with many professional bodies CSI, ISTE, IMS, AMTI, ISIAM, OITS, IACSIT, CSTA, IEEE and IAENG. He was founder secretary of OITS

Rourkela chapter. His current research interests include rough sets, formal concept analysis, knowledge representation, granular computing, data mining and business intelligence.

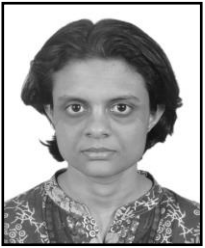

**D. Bhattacharjee** received her B. Tech. degree in computer science and engineering from West Bengal University of Technology, West Bengal, India in 2009. She is an M. Tech (CSE) final year student of VIT University, Vellore, India. She has published two papers in International Conference. She has keen interest in teaching and applied research. Her research interest includes rough computing, granular computing, and soft computing.